\documentclass[a4paper]{article}

\usepackage[english]{babel} \usepackage{hyperref} \usepackage{float}
\usepackage[utf8]{inputenc} \usepackage{amsmath} \usepackage{graphicx}
\usepackage[colorinlistoftodos]{todonotes} \usepackage{tikz}
\usepackage{pdfpages} \usepackage{listings}
\usepackage{afterpage}
\usepackage{lscape}
\usepackage{caption}
\usepackage[a4paper,margin=1in]{geometry} % Márgenes adecuados
\usepackage{authblk} % Para afiliaciones
\usepackage{hyperref} % Para enlaces como ORCID
\usepackage{titlesec} % Mejora títulos y secciones
\usepackage{lmodern} % Fuente moderna y elegante
\usetikzlibrary{arrows,positioning,shapes.geometric}

\titleformat{\section}{\large\bfseries}{\thesection}{1em}{}

\title{From Analog to Digital -- Successful Implementation of IoT Solutions in the Petrochemical Industry}

\author[1]{Noel Portillo\thanks{\href{https://orcid.org/0009-0009-8817-3187}{ORCID: 0009-0009-8817-3187}}}
\affil[1]{\textit{Axxis Technologies}, Houston, TX, USA \\
\href{mailto:noel@axxistechnologies.com}{noel@axxistechnologies.com}}
\affil[2]{\textit{Independent Researcher}, Ahuachapán, El Salvador \\
\href{mailto:noel.portillo@ieee.com}{noel.portillo@ieee.com}}

\date{December 22, 2024}

\begin{document} \maketitle

\section{Abstract}

This document describes the development and implementation of a technological solution based on IoT devices to modernize a machine known as the Cyclone. This equipment is used by a contractor collaborating with petrochemical companies in the state of Texas, performing specialized work in mechanics, engineering, catalytic material replacement, and rescue operations in refinery complexes. The Cyclone machine, with outdated relay logic technology, poses challenges in terms of operational efficiency, critical condition monitoring, and safety.

The project was carried out with the collaboration of specialists in equipment handling, focusing on demonstrating the feasibility of integrating advanced Industry 4.0 technologies into legacy industrial equipment. The methodology included the incorporation of IoT sensors for real-time monitoring, an automated control system, and the digitization of key processes. Preliminary results indicate improvements in the precision of operational control and the ability for remote supervision, highlighting the potential for modernization in critical industrial applications.

This work not only validates the use of IoT devices in obsolete equipment but also sets a precedent for the transition towards more sustainable and efficient technologies in the petrochemical sector.
 
\section{Introduction}

The petrochemical industry, a fundamental pillar of the global economy representing 7\% of the world’s GDP \cite{b1}, faces critical challenges that go beyond its economic impact. From its role as a strategic and geopolitical resource to its significant environmental footprint—contributing 20.3\% of global greenhouse gas emissions in 2019 \cite{b2} — this sector finds itself at a technological crossroads. Despite advances in digitalization and automation characterized by Industry 4.0, the petrochemical industry lags behind in adopting disruptive technologies. This resistance is largely driven by concerns about the reliability, robustness, and cybersecurity of devices, as well as the secrecy of major corporations regarding their processes and industrial know-how.

However, modernization is an unavoidable necessity. The implementation of innovative technologies could not only optimize processes and improve operational safety but also reduce the environmental impact caused by hazardous gas leaks and other incidents. 

This article explores an alternative pathway to drive innovation in the petrochemical industry: focusing on contractor companies. These key players, often operating with outdated equipment and limited budgets, have direct access to industrial complexes and represent fertile ground for experimentation and the development of IoT solutions.

As a case study, this article presents the modernization project of a machine called Cyclone, used by a contractor company based in Houston, TX. This equipment, acquired through auctions and operating with outdated controls, poses challenges for both operational efficiency and worker safety. The project’s objective is to implement a remote monitoring and control system based on an Arduino Opta PLC, enabling safe machine operation and real-time status reporting to the operator.

This approach not only demonstrates how accessible and cost-effective technologies can modernize obsolete equipment but also paves the way for scalable solutions in a sector urgently in need of innovation.
This work showcases two examples of machinery used by the contractor in refinery operations. The first case describes the current operation of the machine known as Screener. The second case involves the development of a functional prototype to control the machine known as Cyclone, also referred to as the "Small Rocket Cyclone."

\section{The Screener Machine}
Let’s take, for example, a piece of equipment called the Screener, primarily designed for sieving and classifying catalytic material. Catalysts are chemical substances in various forms (balls, cubes, irregular shapes) used to facilitate the molecular breakdown of petroleum and obtain byproducts. This equipment performs a process known as screening, which separates the material into different sizes or grades.

\begin{figure}[htbp]
\centerline{\includegraphics{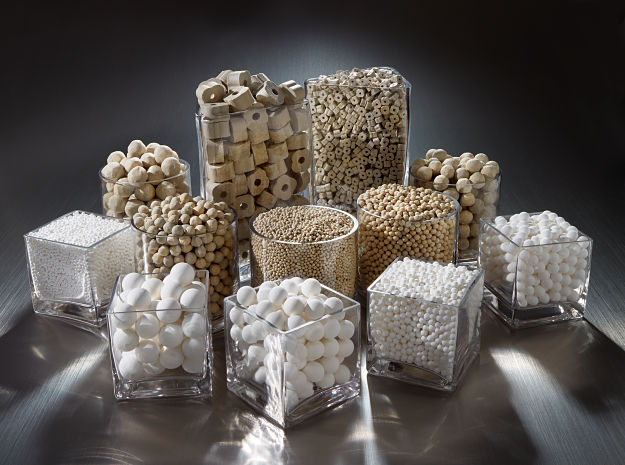}}
\caption{Catalytic Material \cite{b3}}
\label{catalytic}
\end{figure}

The Screener consists of a structure with metallic compartments containing meshes of different sizes. These meshes are arranged in tiers, with openings that decrease in size from top to bottom. The catalytic material is fed into the top section, and through vibrations generated by a motor, the smaller particles pass through successive meshes, while the larger particles are retained in the upper levels. At the end of the process, dust and unusable residues are collected in bags known as "superbags" or, depending on the type of material, in special boxes for storage and further processing. This system efficiently separates usable material from dust and unwanted particles.

From a mechanical perspective, the Screener is equipped with a three-phase motor powered by 480VAC, directly connected to the electrical grid through an AWG 6 cable with four conductors. These cables lead to a Siemens 115D3W box containing a manual switch that the operator must activate to turn on the machine. However, this design presents significant risks. Contact with the metal box during operation can expose the operator to electric shocks if proper precautions are not taken. Additionally, the vibratory motion of the meshes generates dust that can escape through the equipment’s sealing doors. Although the use of personal protective equipment is mandatory, industry statistics reveal cases of petrochemical workers who have developed illnesses related to prolonged exposure to these materials.

The implementation of IoT technologies in equipment such as the Screener could provide significant improvements. For instance, an IoT board could be implemented to control the equipment's startup from a remote station, smart sensors could be installed to monitor motor vibrations, the conditions of the processed material through cameras, and the quality of the surrounding air, thereby reducing risks for operators and optimizing the operational efficiency of the equipment.

\section{The Cyclone Machine}
On the other hand, there are much more specialized pieces of equipment used in higher-risk tasks within petrochemical plant complexes. Another example of such equipment is what we will call "Rocket" due to its resemblance to a rocket. In reality, it is a piece of equipment known in the industry as a Cyclone, which is used in conjunction with a vacuum device (Vacuum) that can either be standalone or installed on a trailer and powered by a John Deere 6068 engine designed for marine and industrial use. These machines are employed for loading and unloading solid catalytic materials.

The mechanical operation of the rocket is based on four pneumatic cylinders, each with a capacity of 100 psi. These cylinders are controlled by two 5-way, double-acting solenoid valves from the manufacturer SMC, specifically from the VQ5000 series. Each valve features two bi-stable solenoids that alternate their operation depending on the action of the other. In other words, when coil A is energized, the airflow is directed in one direction, and when coil B is energized, the airflow shifts to the opposite direction.\footnote{The direction of the airflow is determined by the physical configuration of the hose connections.}

The action of the valves controls the movement of the cylinders, which extend or retract depending on the activation of the solenoid valves. Each pair of cylinders is connected to gates that open and close according to the machine's configuration, which operates as follows:

When the machine is connected to the power supply, a timer relay is activated with an 8-second delay. During this time, the upper gate opens, and the lower gate closes. Then, a timer relay set to 4 seconds closes the upper gate, waits another 8 seconds, and opens the lower gate. Subsequently, another timer relay executes a 4-second delay and closes the lower gate. The initial 8-second timer relay is then reactivated, restarting the process. This cycle continues uninterrupted until the operator turns off the machine.
\vspace*{0.5cm}

\begin{figure}[H]
\centerline{\includegraphics[width=\columnwidth]{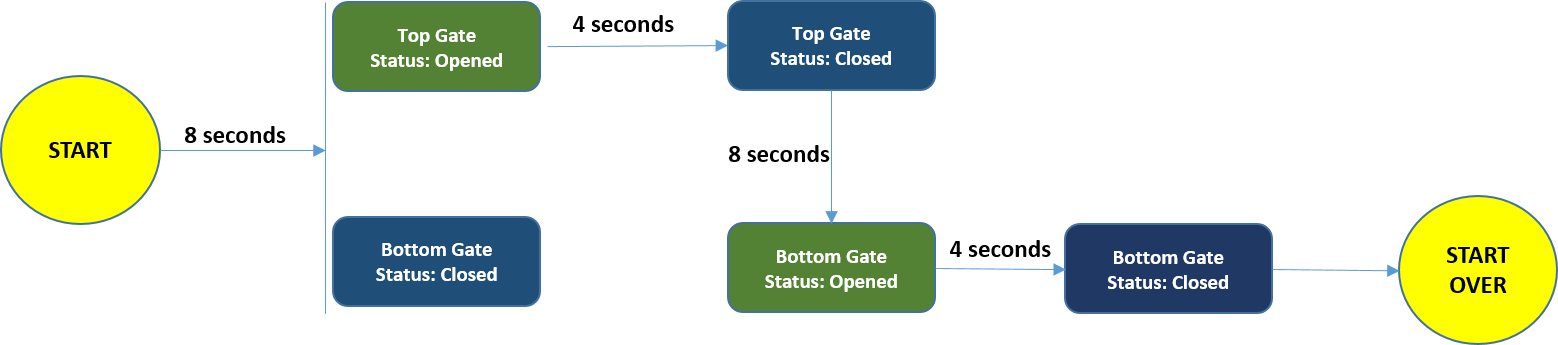}}
\caption{Operational Process of the Cyclone}
\label{process}
\end{figure}

\subsection{Overview of the Electrical Control Circuitry}
The machine's operation is straightforward in design but requires continuous supervision by the operator to ensure optimal performance. This equipment functions in conjunction with a suction system designed to extract chemical materials from reactors, either during catalyst replacement or internal reactor maintenance.

Catalytic materials are highly reactive when exposed to oxygen, which can lead to a significant temperature increase if oxygen enters the main chamber. In certain processes, inert gases are required to displace oxygen from the environment where the material is handled. One of the most commonly used gases for this purpose is hydrogen.

To ensure process safety and prevent undesirable or potentially hazardous reactions, it is essential to keep the upper gate closed while the lower gate is open for material (dust) discharge. This step is critical for maintaining safe operating conditions and preventing accidental oxygen exposure.

The system is based on relay logic and incorporates three Digi-Set timer relays. These timers, configurable via DIP switches, regulate various operations: two control the opening and closing of the upper and lower gates, while the third manages the delay between their activation.

Additionally, two single-pole double-throw (SPDT) relays from Schneider Electric’s Telemecanique series have been integrated to allow manual control of the system via an external interface when needed.

A key safety component of the system is a pair of limit switches that regulate current flow to the timers based on the position of the upper and lower gates. This mechanism ensures that both gates cannot be open simultaneously, mitigating the risk of hazardous exposure.

\begin{figure}[H]
\centerline{\includegraphics[width=\columnwidth]{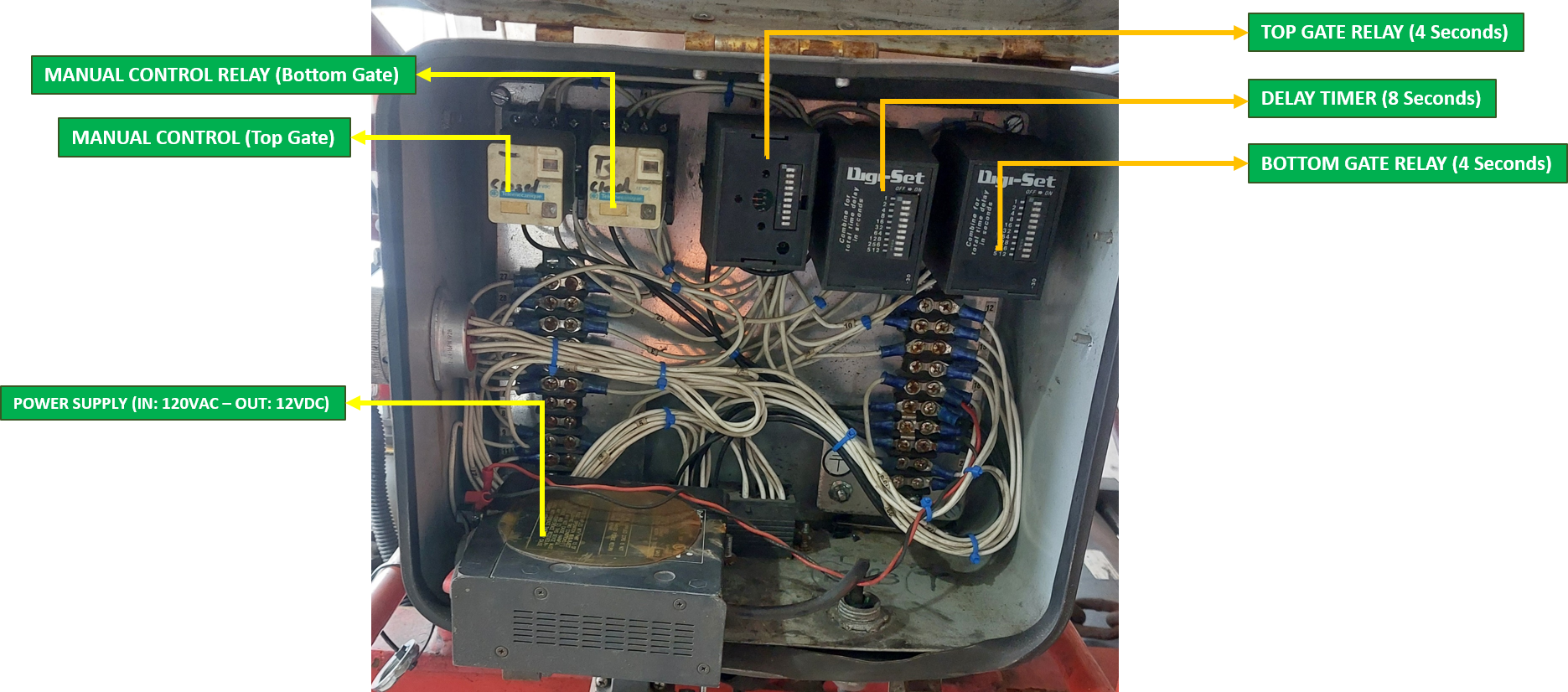}}
\caption{Control Panel of the Cyclone Machine with Key Components Highlighted}
\label{cyclone_original_photo}
\end{figure}

\section{Problem Statement}

The Cyclone machine currently operates using a relay-based control system, which presents several limitations in terms of automation, safety, and monitoring. The absence of real-time data acquisition and remote control capabilities increases operational risks and requires constant manual supervision. Given these challenges, an upgrade to an IoT-based control system is necessary to improve efficiency, enhance monitoring capabilities, and ensure safer operation.

\section{Objectives}

To address these issues, the project aims to implement a modern control system with the following objectives:

\begin{enumerate}
    \item Enable remote control and real-time monitoring of the machine using wireless technology via a portable device.
    \item Integrate temperature sensors at the inlet and outlet of the catalytic material to ensure process stability and prevent overheating.
    \item Incorporate level sensors to automate the opening and closing of the gates, reducing the need for manual intervention and improving operational safety.
\end{enumerate}

\subsection{Hardware Components of the Proposed Solution}

To address the problem, and given that this is a prototype project designed to evaluate the feasibility of using IoT devices in a refinery, the Opta WiFi controller developed by Arduino was utilized. Its most notable features include its development based on Arduino boards, an open-hardware project. This makes it an ideal low-cost solution, as it can also be programmed using the Arduino IDE.

Additionally, for the Opta version, Arduino has developed a programming environment called PLC IDE \cite{b5}, specifically designed to work under the IEC 61131-3 standard. This IDE supports multiple PLC programming languages, such as Ladder Diagram, Functional Block Diagram, Structured Text, Sequential Function Chart, and Instruction List.
In this case, the Arduino Opta AFX00002 hardware version was chosen.

For monitoring the temperature of the Cyclone, two type "K" sensors, model HiLetgo DC 3-5V MAX6675 Module + K Type Thermocouple Temperature \cite{b6}, will be used. For level control, two capacitive sensors, Dinel Serie CLS-23 XiT-30-NTP-N-D-E50-K3 \cite{b7}, will be employed. Regarding the other components, the use of mechanical relays will be eliminated, and the 4 relays integrated into the Arduino Opta will be utilized instead.
Additionally, the limit switches will be reused, and some modifications will be made to the Rocket tank to allow the installation of the temperature and limit sensors.

\subsection{Logical Control Design}

For the design of logical control, Arduino software was used for programming the PLC. As mentioned earlier, one of the advantages of this software is that it is specifically designed for programming Arduino PLC devices \cite{b8}. However, during its use, we identified certain deficiencies that the manufacturer could improve upon \cite{b9}. Despite this, the use of this software is not indispensable, as there are other alternatives that can meet the stated objectives.

The process began with writing a \textit{Sketch}, which is the term used in Arduino for programs written in C++ for these devices. This \textit{Sketch} included the control of the MAX6675 boards, whose integration is essential because type K thermocouples generate a very low voltage signal (in the range of microvolts to millivolts), depending on the temperature difference between the hot junction (measurement point) and the cold junction (reference) \cite{b10}. The MAX6675 acts as a precision amplifier that scales this signal, allowing for accurate data readings \cite{b11}.

Additionally, the initial state of the machine was established, defined as follows:
\begin{itemize}
    \item Upper Gate: Closed.
    \item Lower Gate: Closed.
\end{itemize}

Since the valves used are three-position and bi-stable (built with two solenoids, A and B), it is important to note that these solenoids maintain their position until the other one is energized \cite{b12}. For this reason, in the sequence of events, it is essential to ensure that the solenoid not in use is de-energized. That is, if solenoid A is energized, solenoid B must be de-energized, and vice versa, to ensure the proper functioning of the system.

{\subsection{Arduino Sketch for MAX6675 Sensors}}

\begin{lstlisting}[caption={Arduino Sketch for Reading MAX6675 Sensors}, label={lst:max6675}]
#include <max6675.h>
// Shared Pins
const int SCK = 5;
const int SO = 6; 

// CS Pins
const int CS1 = 0;   // Chip Select for Upper Sensor
const int CS2 = 1;   // Chip Select for Lower Sensor

// Sensor Initialization...
MAX6675 thermocouple1(SCK, CS1, SO);
MAX6675 thermocouple2(SCK, CS2, SO);

void setup() {
    // Initial State
    Upper_Gate_Open     =  FALSE;
    Upper_Gate_Close    =  TRUE;
    Lower_Gate_Open     =  FALSE;
    Lower_Gate_Close    =  TRUE;
    Serial.begin(9600);
    Serial.println("Initializing Sensors...");
}

void loop() {
    // Upper Temperature Sensor Read
    Temperature_Sensor_Upper = thermocouple1.readCelsius();
    Serial.print("Upper Sensor: ");
    Serial.println(Temperature_Sensor_Upper);

    // Lower Temperature Sensor Read
    Temperature_Sensor_Lower = thermocouple2.readCelsius();
    Serial.print("Lower Sensor: ");
    Serial.println(Temperature_Sensor_Upper);
    delay(1000); // Read Every Second
}
\end{lstlisting}

\subsection{Ladder Solution Implementation}
Next, we proceed to write our program in Ladder. It is important to highlight that Arduino software offers certain advantages compared to other manufacturers. This software allows interaction between structured programming languages and Ladder logic through the use of shared global variables. This is highly desirable in development environments, as programming in C++ can be used to easily interact with and modify the PLC's behavior \cite{b13}.

Although platforms like Codesys or TwinCAT also offer this functionality, they sometimes require more specific configurations, making the simplicity of the Arduino environment an advantage for prototyping and rapid development projects \cite{b14}.

Furthermore, Arduino software facilitates integration with microcontrollers that use libraries written in C++ and, with certain configurations, even in Python. This significantly expands the possibilities when developing innovative and customized solutions, enabling the combination of industrial control capabilities with the flexibility of microcontrollers and modern technologies \cite{b15}.

However, it is important to note that, in advanced industrial systems, other manufacturers may offer more robust and specific tools for certain critical applications, which should be considered depending on the project's scope and requirements \cite{b16}.

\section{Ladder Logic Diagram}

\begin{figure}[H]
\centerline{\includegraphics[width=\columnwidth]{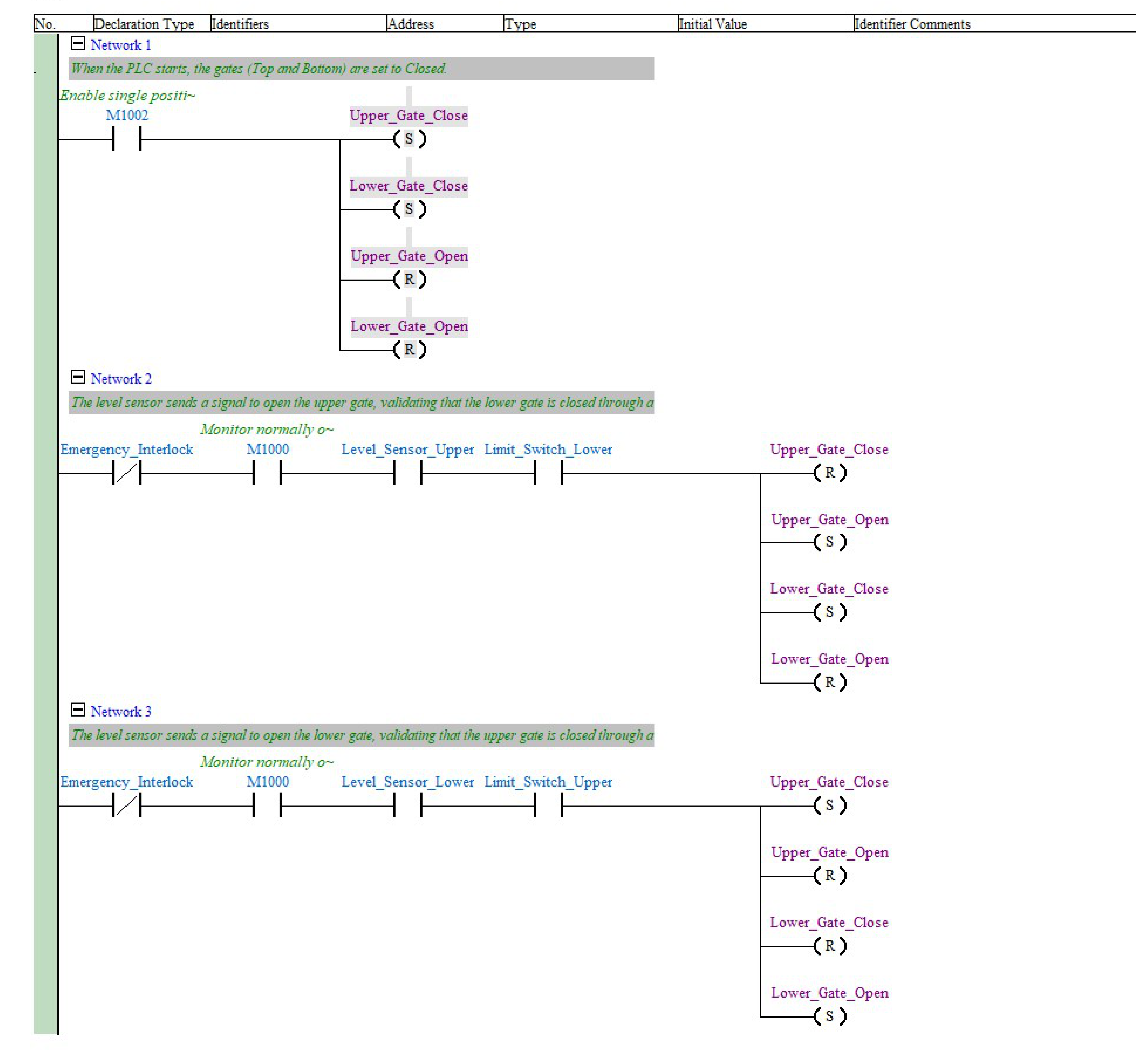}}
\end{figure}

\begin{figure}[H]
\centerline{\includegraphics[width=\columnwidth]{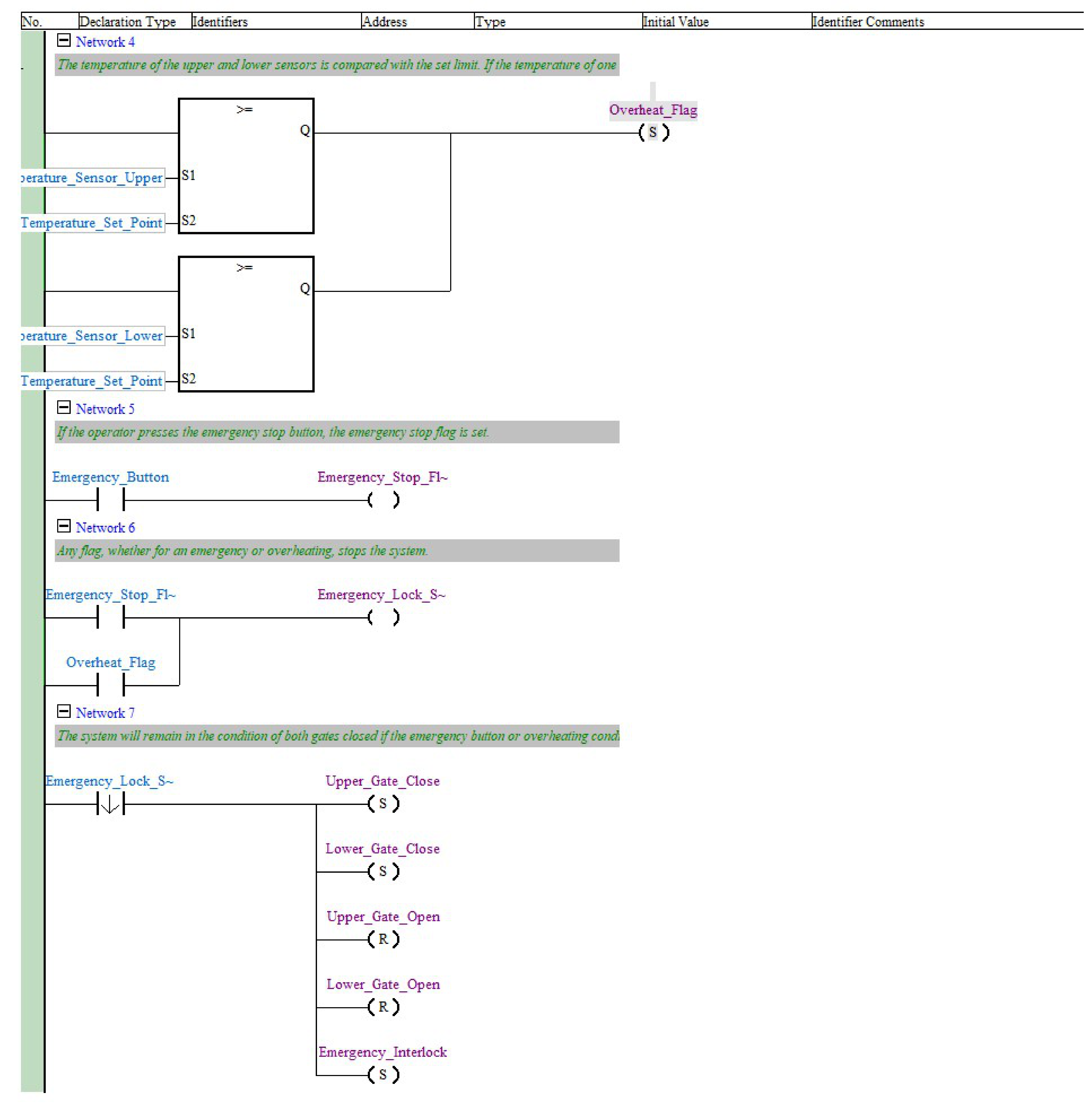}}
\end{figure}

\section{Analysis of Results}

The design and implementation of the prototype took approximately a month and a half, excluding the time required for parts acquisition and the reconfiguration of the Cyclone tank. Metallurgical fabrication work was necessary to drill holes and position the sensors.

The main issue encountered during testing was that the inductive sensors used to measure the tank's fill limits experienced interference due to suspended material and the flow of material falling onto the sensor. To address this issue, we consulted the sensor installation manual, where we found the solution. A metal plate was installed above the sensor to prevent direct contact with the falling material, implemented as follows:

\begin{figure}[H]
\centerline{\includegraphics[width=\columnwidth]{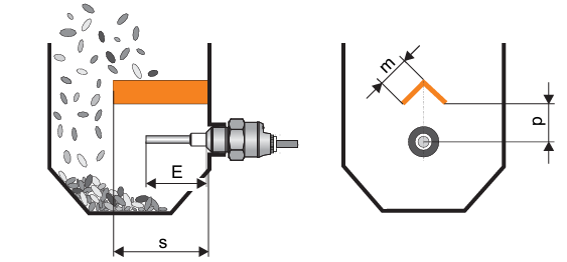}}
\captionof{figure}{Protective metal plate for horizontal sensor installation\cite{b17}}
\label{level_sensor}
\end{figure}

The following equations describe the parameter relationships as a function of the electrode length:

\begin{flalign*}
\hspace{0pt} s &\approx \frac{4}{3} E &&\\
\hspace{0pt} p &\approx \frac{3}{4} E &&\\
\hspace{0pt} m &\approx \frac{2}{3} E &&
\end{flalign*}

where \( E \) represents the electrode length in millimeters.

In the case of the model used, Dinel CLS-23 XiT-30-NTP-N-D-E50-K3, the electrode measures 50mm. Therefore, the dimensions for the protective plate are:

\begin{flalign*}
\hspace{0pt} S &= \frac{4}{3} E  &&\\
\hspace{0pt} S &= \frac{4}{3} (50\text{mm})  &&\\
\hspace{0pt} S &= 66.67\text{mm}  &&\\
\hspace{0pt} P &= \frac{3}{4} E  &&\\
\hspace{0pt} P &= \frac{3}{4} (50\text{mm})  &&\\
\hspace{0pt} P &= 37.5\text{mm}  &&\\
\hspace{0pt} M &= \frac{2}{3} E  &&\\
\hspace{0pt} M &= \frac{2}{3} (50\text{mm})  &&\\
\hspace{0pt} M &= 33.33\text{mm}  &&
\end{flalign*}

Once this issue was resolved, controlled tests were conducted in the workshop, yielding satisfactory results. After two weeks of controlled testing, the equipment was finally ready to be tested in a real environment.

\begin{figure}[H]
\centerline{\includegraphics[width=\columnwidth]{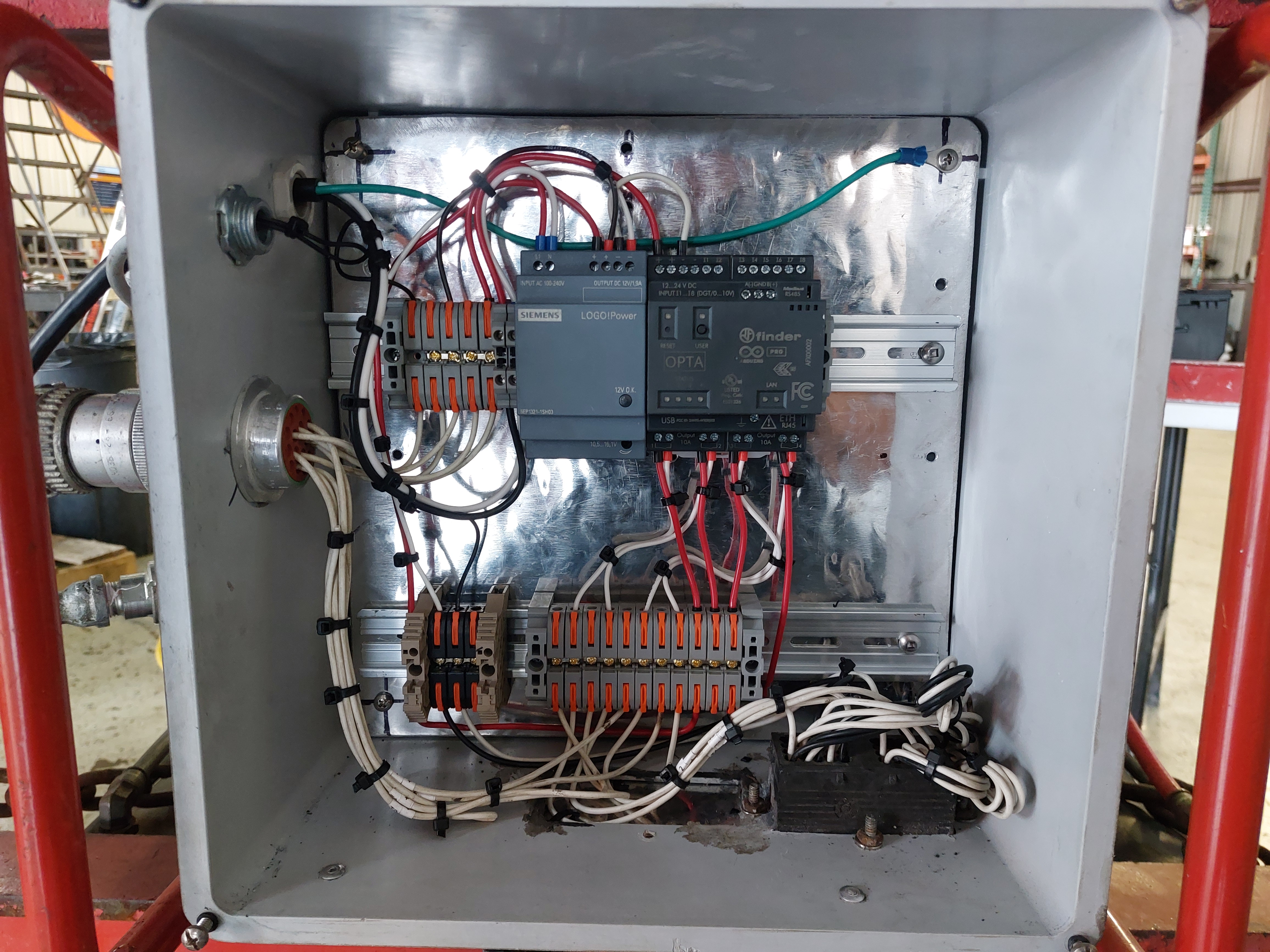}}
\captionof{figure}{Assembly of the automatic control prototype designed for the automation and monitoring of the cyclone.}
\label{prototype_box}
\end{figure}

\section{Conclusions}

The cyclone was sent to a refinery for two weeks, where it was used in the replacement of the catalytic material in a reactor. No significant incidents were reported regarding the equipment's operation, except for a damaged pneumatic cylinder due to improper tightening of the mounting screws. This incident was entirely unrelated to the new control solution implemented in the equipment.

Additionally, positive feedback was received from the operators, who highlighted the system's safety, convenience, and ease of operation.

As a result of the success achieved with this prototype, further research has been requested to integrate this equipment with other systems used in conjunction, such as the vacuum equipment. Since the process requires synchronization between the vacuum system and the opening and closing of the gates, this integration would be a key step. Furthermore, it has been suggested to incorporate oxygen and pressure sensors at the air or hydrogen inlet of the cylinders to ensure a safer and more optimal operation.

\newpage 
\section{Appendix} 

\begin{figure}[H]
\centerline{\includegraphics[width=\columnwidth]{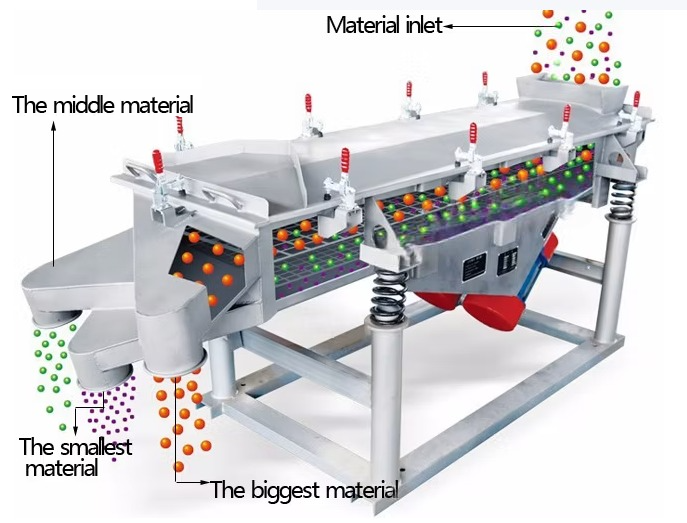}}
\caption{Screener Machine \cite{b4}}
\label{screener}
\end{figure}

\begin{figure}[H]
\centerline{\includegraphics[width=\columnwidth, angle=-90]{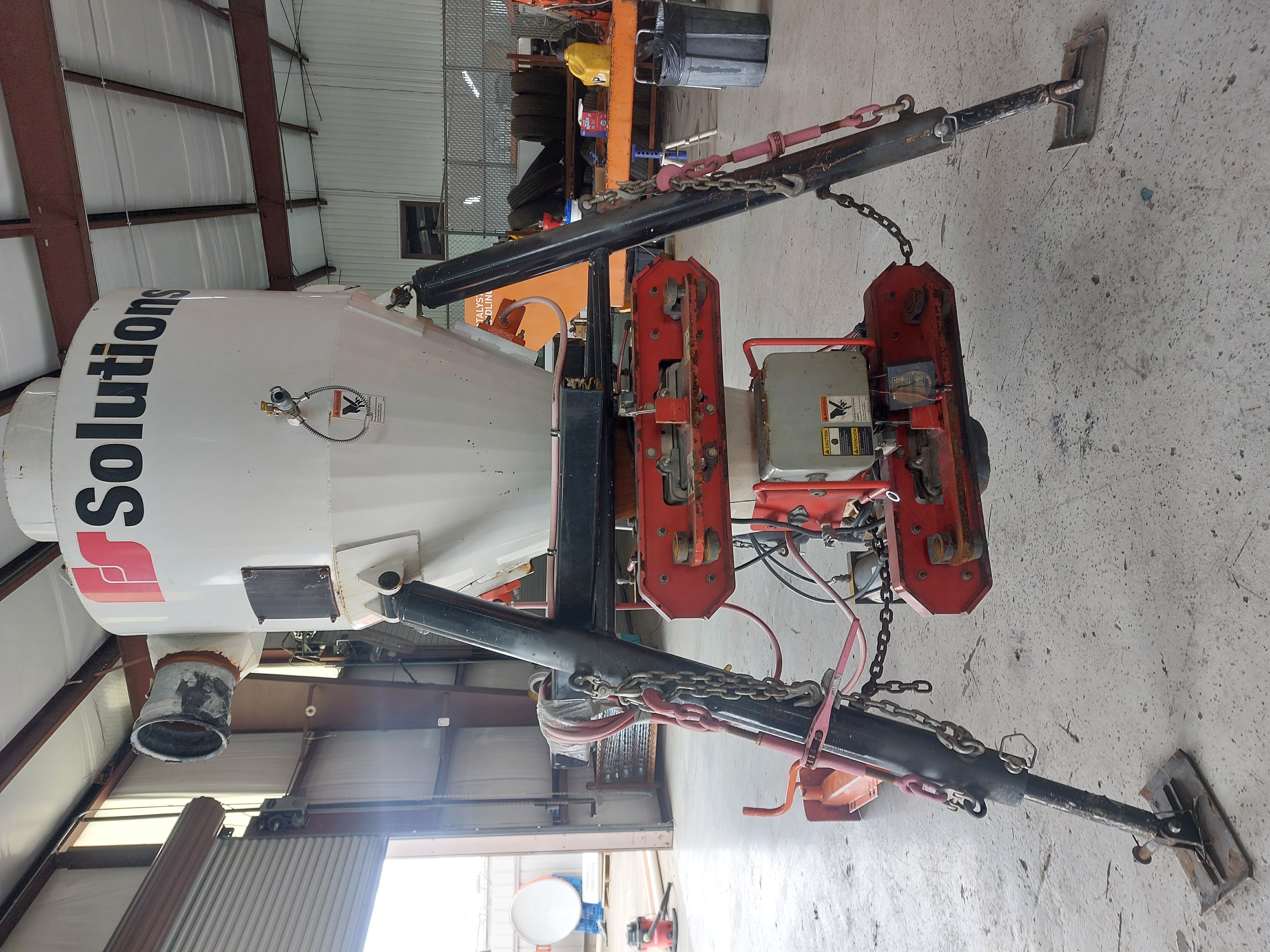}}
\caption{Cyclone Machine --- Know as Rocket}
\label{cyclone}
\end{figure}

\begin{figure}[H]
\centerline{\includegraphics[width=\columnwidth]{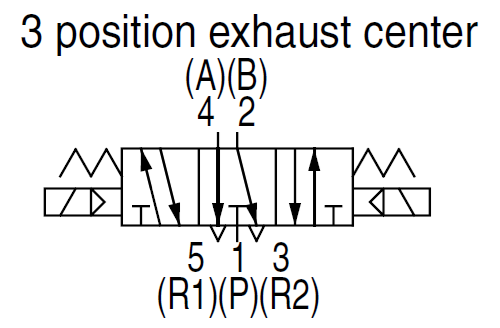}}
\caption{Pneumatic diagram of the SMC VQ5000 5-way, two-position solenoid valve}
\label{vq5000}
\end{figure}

\begin{figure}[H]
\centerline{\includegraphics[width=\columnwidth]{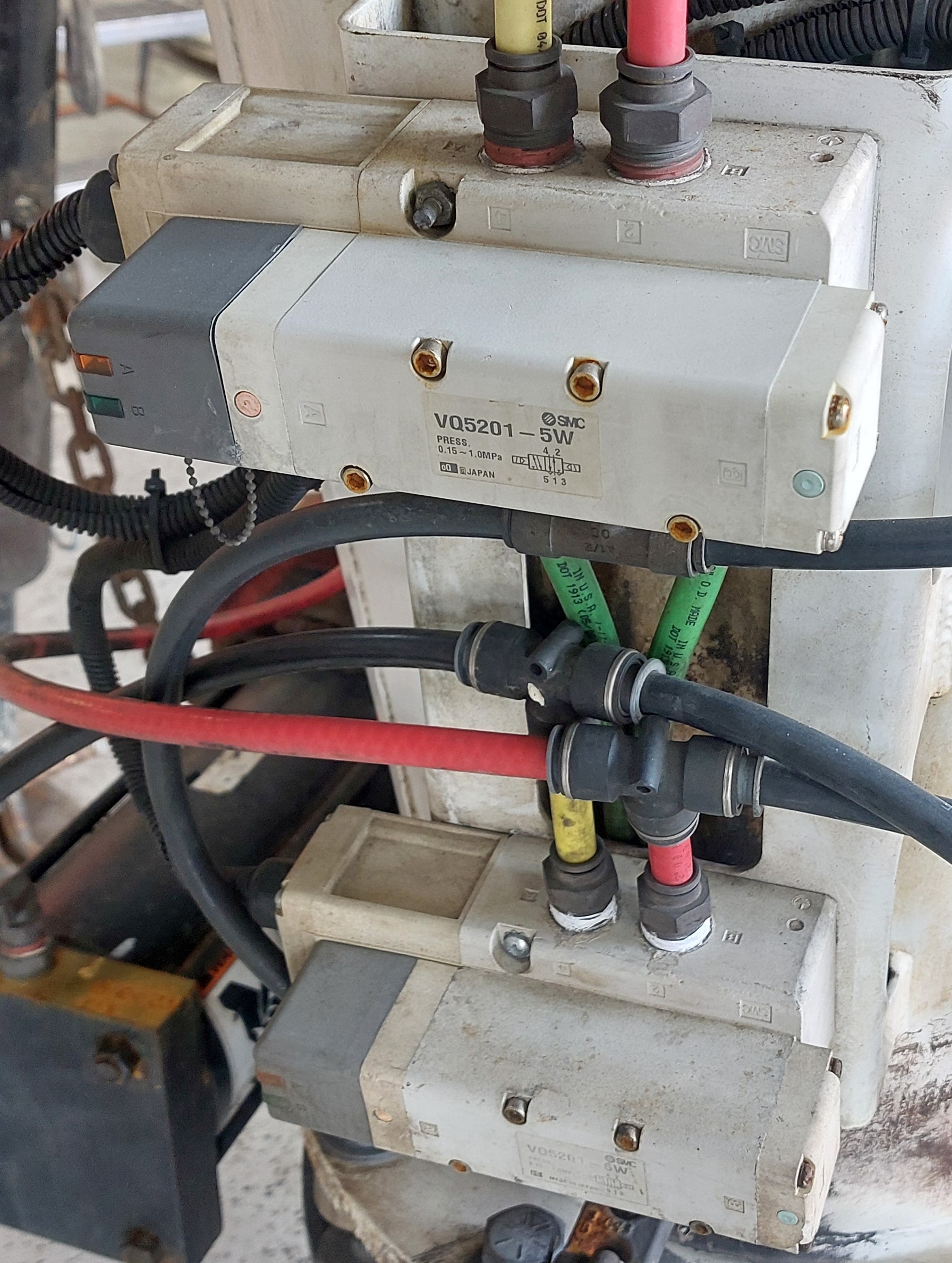}}
\caption{Actual photo of the SMC VQ5000 series valves}
\label{vq5000-real}
\end{figure}

\begin{figure}[H]
    \centering
    \rotatebox{90}{
        \includegraphics[width=\paperwidth,height=\paperheight,keepaspectratio]{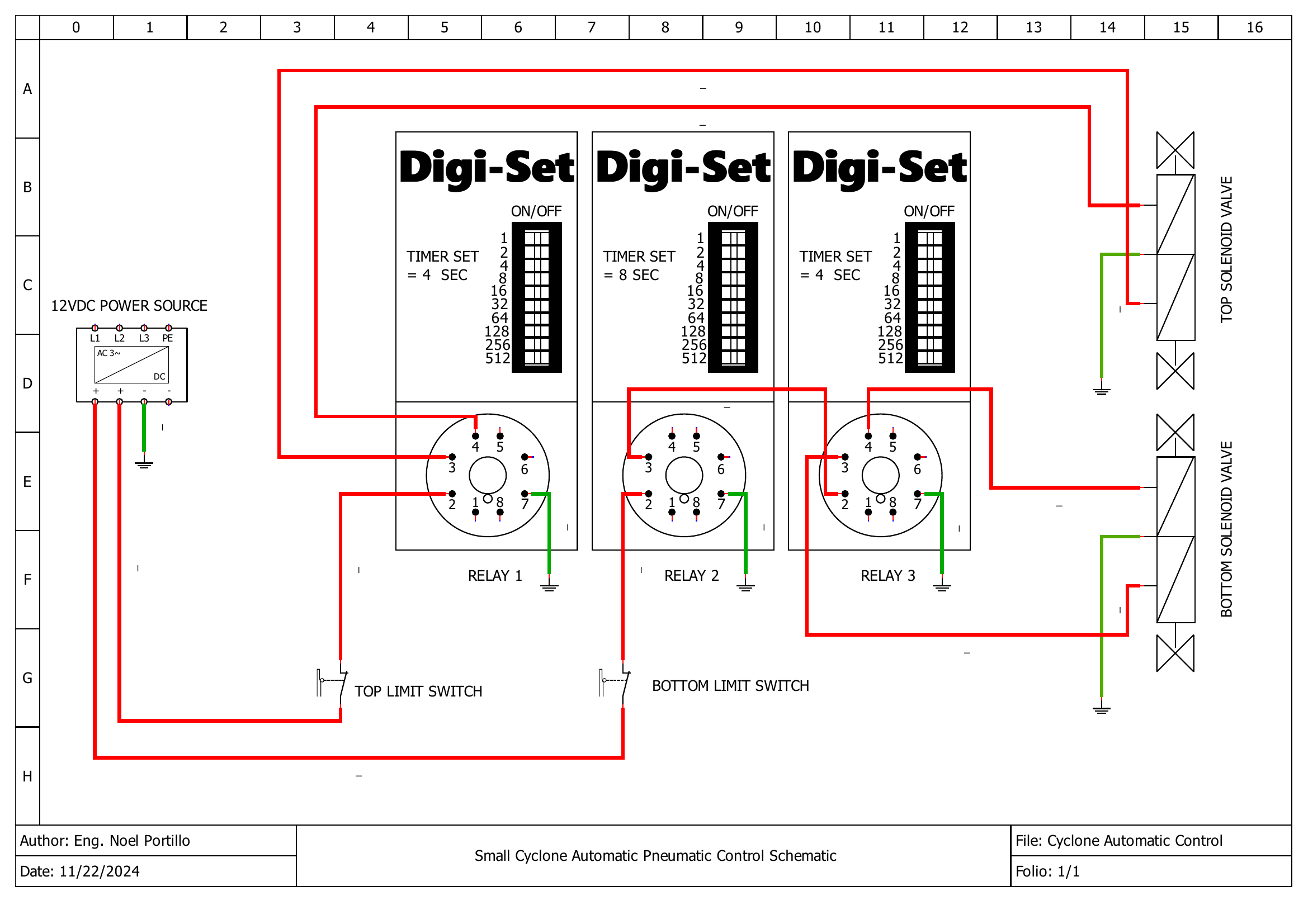}
    }
    \caption{Original Electrical Control Circuit Diagram of the Cyclone}
    \label{cyclone_original_schematic}
\end{figure}

\clearpage
\begin{figure}[p]
    \centering
    \includegraphics[width=\textwidth]{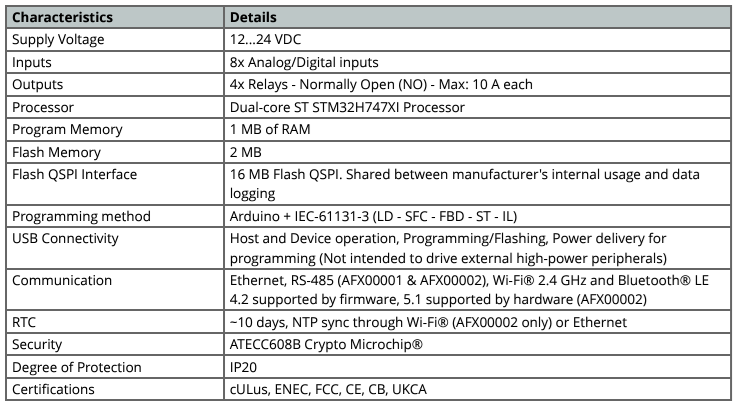}
    \caption{Hardware Specifications of the Arduino Opta AFX00002 IoT Controller}
    \label{AFX00002}
\end{figure}

\clearpage
\begin{figure}[p]
    \centering
    \rotatebox{90}{
        \includegraphics[width=\paperwidth,height=\paperheight,keepaspectratio]{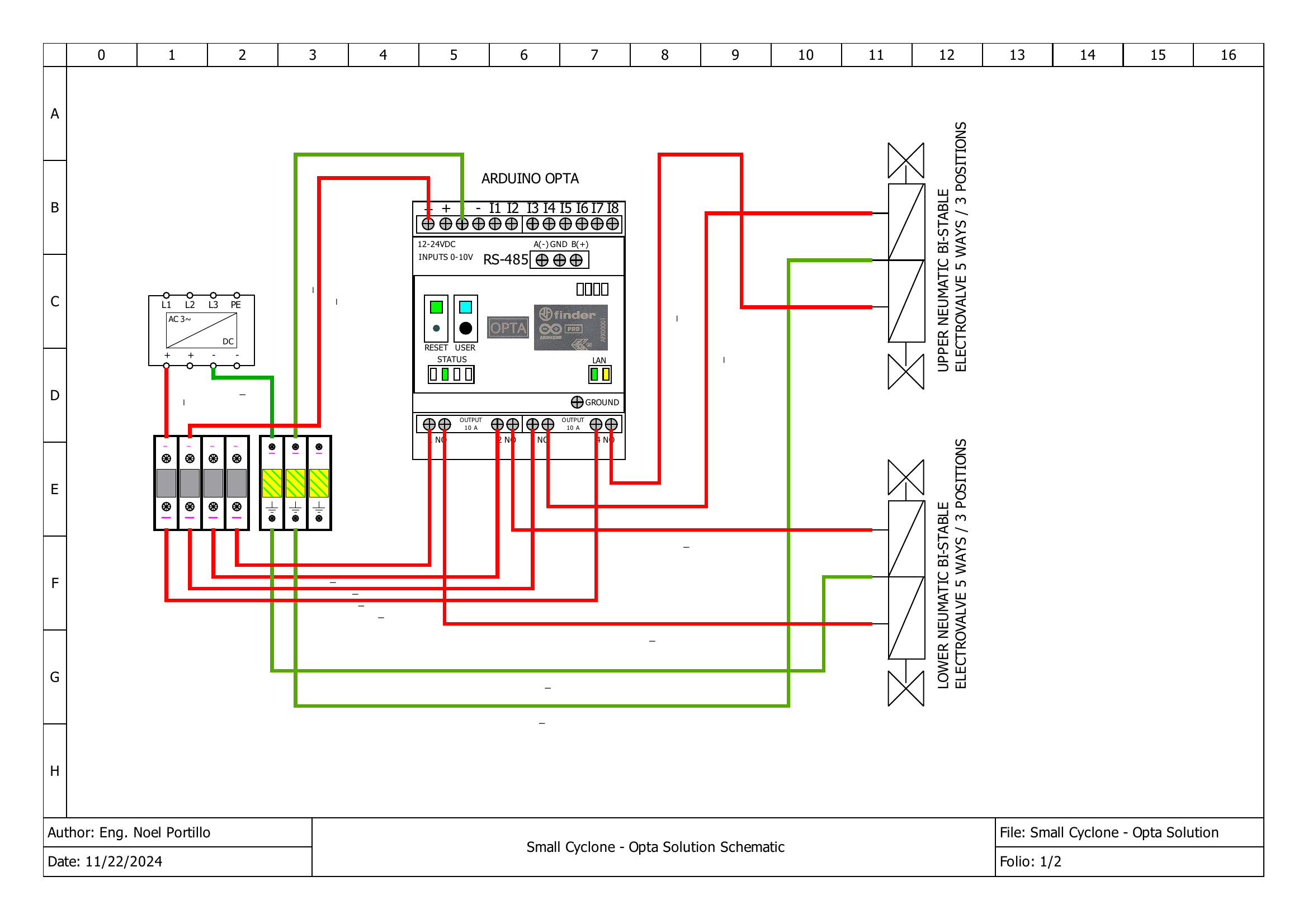}
    }
    \caption{Proposed Electrical Control Circuit Diagram of the Cyclone (1)}
    \label{cyclone_proposed_schematic_1}
\end{figure}

\clearpage
\begin{figure}[p]
    \centering
    \rotatebox{90}{
        \includegraphics[width=\paperwidth,height=\paperheight,keepaspectratio]{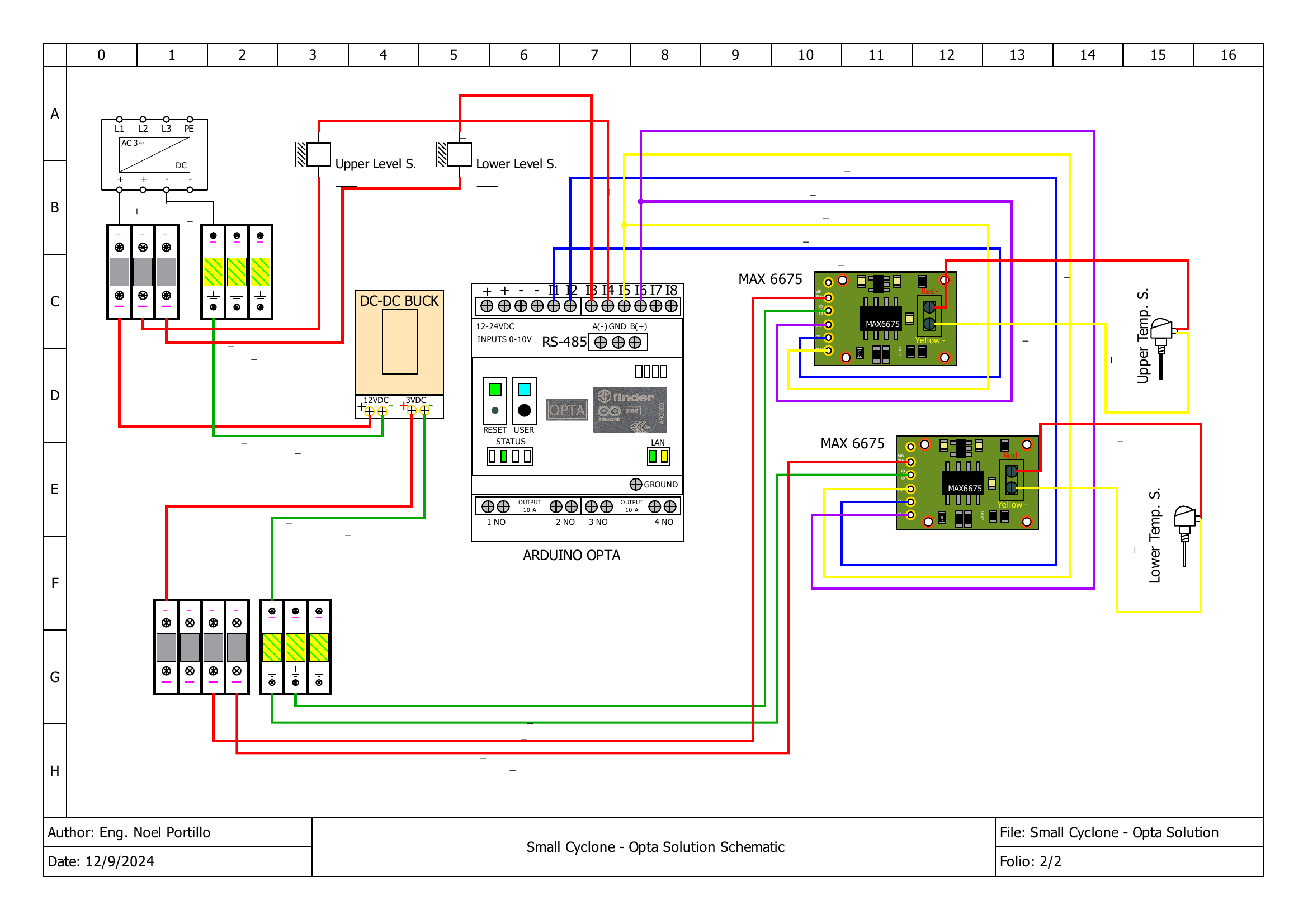}
    }
    \caption{Proposed Electrical Control Circuit Diagram of the Cyclone (2)}
    \label{cyclone_proposed_schematic_2}
\end{figure}


\begin{thebibliography}{00}
\bibitem{b1} Chemical Industry Journal, "£5.7 trillion contribution," 2024. [Online]. Available: \url{https://www.chemicalindustryjournal.co.uk/5-7-trillion-contribution}. [Accessed: 28-Jan-2025].
\bibitem{b2} Parlamento Europeo, "Emisiones de gases de efecto invernadero por país y sector – Infografía," 2018. [Online]. Available: \url{https://www.europarl.europa.eu/topics/es/article/20180301STO98928/emisiones-de-gases-de-efecto-invernadero-por-pais-y-sector-infografia}. [Accessed: 28-Jan-2025].
\bibitem{b3}Ingeniería Química Reviews, "Catalizadores sólidos en reactores químicos," 2021. [Online]. Available: \url{https://www.ingenieriaquimicareviews.com/2021/08/catalizadores-solidos-reactores-quimicos.html}. [Accessed: 28-Jan-2025].
\bibitem{b4} Sanyuan Tang, "Made-in-China," [Online]. Available: \url{https://xxsanyuantang.en.made-in-china.com/}. [Accessed: 28-Jan-2025].
\bibitem{b5} Arduino, "Arduino PLC IDE," [Online]. Available: \url{https://www.arduino.cc/pro/software-plc-ide/}. [Accessed: 28-Jan-2025].
\bibitem{b6} HiLetgo, "Product Detail," [Online]. Available: \url{http://www.hiletgo.com/ProductDetail/1953467.html} [Accessed: 28-Jan-2025].
\bibitem{b7} Dinel, "Capacitive level switches CLS-23 – Data Sheet," 2024. [Online]. Available: \url{https://tinyurl.com/yf3v385n} [Accessed: 28-Jan-2025].
\bibitem{b8} Arduino, "Arduino PLC IDE," Arduino Official Documentation, 2023. [Online]. Available: \url{https://www.arduino.cc/}. [Accessed: 28-Jan-2025].
\bibitem{b9} R. Struharik, "Arduino Applications in Industrial Automation," Control.com, 2023. [Online]. Available: \url{https://control.com/technical-articles/arduino-applications-in-industrial-automation/}. [Accessed: 28-Jan-2025]..
\bibitem{b10} MAX6675 Datasheet, "K-Type Thermocouple Amplifier," Maxim Integrated, 2021. [Online]. Available: \url{https://www.maximintegrated.com/}. [Accessed: 28-Jan-2025]..
\bibitem{b11} Thermocouple Info, "Thermocouple Information Resource," 2023. [Online]. Available: \url{https://www.thermocoupleinfo.com/}. [Accessed: 28-Jan-2025].
\bibitem{b12} Festo, "Technical Report: TR-300003-EN-Ver2," 2023. [Online]. Available: \url{https://www.festo.com/net/en-in_in/SupportPortal/Downloads/265621/432404/TR-300003-EN-Ver2.pdf}. [Accessed: 28-Jan-2025].
\bibitem{b13} Arduino Official Documentation, "Programming Arduino PLC IDE," 2023. [Online]. Available: \url{https://www.arduino.cc/}.
\bibitem{b14} Codesys, "The IEC 61131-3 Automation Software," Codesys Documentation, 2023. [Online]. Available: \url{https://www.codesys.com/}.
\bibitem{b15} P. Anderson, "Control of Microcontrollers in Industrial Automation," Industrial Control Systems, vol. 15, no. 4, pp. 78-85, 2021.
\bibitem{b16} Beckhoff, "TwinCAT System Overview," Beckhoff Automation, 2023. [Online]. Available: \url{https://www.beckhoff.com/}.
\bibitem{b17} Dinel, s.r.o., "Capacitive level sensors CLS–23: Instruction manual," n.d. [Online]. Available: \url{https://tinyurl.com/yc2kbsun}. [Accessed: 31-Jan-2025].
\end{thebibliography}
\end{document}